# **Title** 10 Simple Rules for Improving Your Standardized Fields and Terms

## Authors

Rhiannon Cameron (1,2), Emma Griffiths (1,3), Damion Dooley (1,4), William Hsiao (1,5)
Corresponding author: wwhsiao@sfu.ca

#### Affiliations

1 Centre for Infectious Disease Genomics and One Health, Faculty of Health Sciences, Simon Fraser University, Burnaby, BC, Canada

#### Acknowledgements

Charlotte Barclay (1,6)

#### ORCiD

2 https://orcid.org/0000-0002-9578-0788
3 https://orcid.org/0000-0002-1107-9135
4 https://orcid.org/0000-0002-8844-9165
5 https://orcid.org/0000-0002-1342-4043
6 https://orcid.org/0000-0001-8008-8249

#### Author Contributions

Conceptualization by EG and RC. Manuscript writing by RC. Revisions and Editing by RC, EG, DD, and WH.

## Abstract

Contextual metadata is the unsung hero of research data. When done right, standardized and structured vocabularies make your data findable, shareable, and reusable. When done wrong, they turn a well intended effort into data cleanup and curation nightmares. In this paper we tackle the surprisingly tricky process of vocabulary standardization with a mix of practical advice and grounded examples. Drawing from real-world experience in contextual data harmonization, we highlight common challenges (e.g., semantic noise and concept bombs) and provide actionable strategies to address them. Our rules emphasize alignment with Findability, Accessibility, Interoperability, and Reusability (FAIR) principles while remaining adaptable to evolving user and research needs. Whether you are curating datasets, designing a schema, or contributing to a standards body, these rules aim to help you create metadata that is not only technically sound but also meaningful to users.

## Funding

This work was funded by Canadian COVID Genomics Network (CanCOGeN) VirusSeq Project (Genome Canada grant number E09CMA) and by Genome Canada and Genome BC Computational Biology and Bioinformatics Grant (project number 286GET) to WH as well as the Canadian Institutes of Health Research Canadian Graduate Scholarship for Doctoral students (funding reference number 187519) to RC. WH was supported by the Michael Smith Health Research BC Scholar Award.

## Introduction

In the era of genomic pathogen and antimicrobial resistance (AMR) surveillance, high-quality sequence and contextual data (i.e., metadata) have been vital to our ability to understand and respond to outbreaks. While genetic information is one important piece of public health investigations, the associated contextual data informs the meaning of both qualitative and quantitative analyses, facilitating an informed understanding of what is being analyzed and what the results signify. By contextual data we mean the epidemiological, clinical, environmental, and laboratory information describing samples, hosts, measurements, parameters, conditions, and methods that allow us to make sense of the sequence data. However, it is often collected and stored using organization-specific terminology, or as free text in a loosely defined, semi-structured manner. This lack of standardization makes data highly variable in both content and structure as different projects, programs, and agencies prioritize different research questions; apply different methodologies and instruments; and must address the various submission requirements of the platforms in which the data flows (e.g., data collection devices, subject-area specific databases, regional vs national databases, and public repositories). Standardization forms a basis of comparison and commonality that allows for an informed interpretation of what is being analyzed and the resulting conclusions.

The COVID-19 pandemic highlighted the challenges of consolidating heterogeneous data and the importance of data harmonization to improve interoperability, reusability, quality, and efficiency[1,2]. Variability in datasets increases uncertainty for data users, especially when data definitions are unavailable or of poor quality. It is now becoming widely recognized that harmonizing data, by applying agreed-upon vocabularies and standardized formats, reduces these uncertainties. Consequently, laboratories worldwide are prioritizing the development and implementation of data standards as genomic surveillance programs expand to a wider array of pathogens, hosts, and environments, from healthcare-associated pathogens to vector-borne pathogens to wastewater surveillance. While genomics and bioinformatics capacity building and training programs are increasingly available, expert knowledge of best practices and principles for developing data standards is often localized, trapped even, within communities of practice. This knowledge needs to be made accessible to public health and research institutions to facilitate the development of interoperable specifications that improve data utility and sharing within and across institutions.

An open community of practice that is well established in semantic data science and facilitates the development of interoperable data specifications in biomedical research, but less known in the public health community, is the Open Biological and Biomedical Ontology (OBO) Foundry[3]. Ontologies are sets of controlled (standardized) vocabulary that are structured in hierarchies using logical relationships that enable more complex queries and linked data (knowledge webs), as well as different kinds of classifications for analyses. Ontologies are developed by domain area experts and maintained through consensus for, and by, the community. As such, the meanings of terms are meant to be universal, rather than institutional- or project-specific, and are disambiguated using universal identifiers (e.g., Uniform Resource Identifiers (URIs)). Ontology term labels and synonyms take on a secondary, more cosmetic role intended for human comprehension, leaving term URIs and definitions with the main work of determining comparability of data items, allowing a "one-size fits all" terminology or nomenclature system which contributes to interoperability. Communities of practice like the OBO Foundry articulate and implement best principles and practices to enable the reuse of terminology across domains and sectors. Furthermore, there are different community registries and portals enabling Findability, Accessibility, Interoperability, and Reusability (FAIR)[4] ontology development and implementation (e.g., EBI OLS[5], Ontobee[6], BioPortal[7]), as well as data modeling languages and tools for improving reuse and interoperability (e.g., Protégé[8], LinkML[9], ROBOT[10], OntoFox[11]). While the OBO Foundry primarily focuses on describing information in academic biomedical research, as the demand for data harmonization in public and food safety healthcare systems has increased, the OBO Foundry community has been adapting and expanding ontologies to better address both academic and non-academic contextual data challenges.

At the Centre for Infectious Disease Genomics and One Health (CIDGOH), we collaborate with different laboratories, organizations and genomics initiatives to develop data specifications, as well as information capture and communication templates and applications. We have combined OBO Foundry ontologies, tools, and best practices with the engagement of data generators and consumers in public health, food safety, One Health and AMR surveillance to develop interoperable data standards based on a common, extensible framework[2,12]. Our team developed the Canadian COVID-19 Genomics Network (CanCOGeN) VirusSeq specification[1,13], the Canadian Genomics Research and Development Initiative for Antimicrobial Resistance (GRDI-AMR) specification[14,15], the Alberta Microbiota Repository (AMBR) specification[16], Canadian and International MPox specifications[17], the Public Health Alliance for Genomic Epidemiology (PHA4GE) COVID-19 specification[2,18], as well as the PHA4GE Wastewater Surveillance (WWS)[19], hAMRonization[20], and Quality Control (QC) tags[18], Highly Pathogenic Avian Influenza (HPAI)[21] specifications. This work has also led to many practical lessons learned regarding data standardization that can contribute to best practices. In this document, we distill these lessons into 10 simple rules for standards developers to improve consistency, reusability and interoperability of data structures meant for contextual data capture and sharing in laboratories around the world.

## Contextual Data Challenges

Our work in curating and harmonizing various types of pathogen genomics contextual data, in addition to engagement with other groups working on the same task, resulted in the identification of recurring issues that posed challenges for data users and standards developers[1]. Challenges we have identified in contextual data application and harmonization can be summarized as one or more of the following:

**Non-Harmonized Vocabulary:** Vocabulary that has meaningful differences across fields, terms, and formats that make it difficult to combine with other datasets that would otherwise be compatible and comparable.

**Semantic Noise**: Semantic noise is the confusion or miscommunication caused by semantic ambiguity, which occurs when the same words have different meanings in different contexts or when different words are interpreted as having the same meaning.

**Missing Vocabulary:** The absence of appropriate controlled vocabulary during data collection can lead to users not applying controlled vocabularies out of frustration, or applying controlled vocabulary incorrectly in order to force their data into an ill-fitting template.

**Concept/Data Structure Incompatibilities:** Data standards can suffer from conceptual and structural differences, such as using different classification schemes and levels of specificity in terms/fields/categories, resulting in imperfect matches when standards are mapped across each other. This often occurs because many standards are developed in silos, with specific use cases and data needs in mind using differing or non-standard practices.

**Word Bombs:** Word bombs occur when a proper noun can be modified by many different descriptors leading to picklists of seemingly endless combinations (e.g., farm; fish farm; turkey farm; chicken farm; crop farm; chicken and crop farm; chicken, turkey and crop farm, *ad infinitum*).

**Concept Bombs:** Concept bombs occur when a field or term is packed with many modifiers, resulting in terminology that is overly specific and difficult to reuse/map (e.g., “previous SARS-CoV-2 infection in the last 6 months with treatment"), causing data standard bloat, and would be be better served by separation into individual fields representing the different concepts.

**Timeline Terms:** Data descriptors intended to capture information at relative points in time (e.g., "most recent test date") can quickly become wrong or obsolete as time passes and additional information is collected.

**Ontologies vs. Implementations:** Ontologies tend to have very specific, unambiguous (and sometimes wordy) labels - which is something we advocate for within the rules. However, this can sometimes create barriers to utilizing ontologies when users are reluctant to relinquish the terms they are accustomed to.

**Ontology Term Branch Patterns:** New data standards developers may find it challenging to locate ontology terms for reuse due to the structure of the ontologies. OBO Foundry ontologies structure all knowledge according to an upper level hierarchy - specifically, everything can be represented as a material entity (thing), a process, a quality (property or characteristic of a thing), or a datum about a thing or process. Different entities/processes can be found in different thematic ontologies (e.g. Food Ontology, Anatomy Ontology, Environment Ontology), and may also be found in different branches or under different parent classes within the same ontology, depending on their nature. Design patterns in ontologies may mean that not all domain knowledge is co-located in the same place (e.g. standardized vocabulary needed for genomic surveillance of food production facilities may be found across ENVO, FoodOn, GenEpiO, OBI, UBERON, etc.).

**Entity vs. Its Usage:** Sometimes what a thing was built to do is not what it is used for (e.g., a milk crate's intended use is transporting milk, but milk crates are often used in university dorms as bookshelves and stepping stools). Understanding the true universality of the information you are representing is critical for defining and using standardized terms to model processes and capture experimental methods (e.g., a layer chicken is a chicken that lays eggs; however, a layer chicken can also be used as meat in different products when it is incapable of providing eggs any longer). Inexperienced data standard developers sometimes do not look beyond their own use case or context in defining or creating new vocabulary which limits its use/reuse.

**Entities vs. Information "About" Entities:** One of the benefits of using ontologies in data standards is that they are both human- and machine-readable resources. Machine-readability requires a greater degree of language precision than our human communication. To facilitate this, ontologies distinguish between entities and information about entities - known as information content entities (e.g., an "host" vs the information content entities "host health state" and "host age"). Ontologies also differentiate between entities and data about an entity (e.g., "amino acid mutation", An ontology conceptual term for understanding what an amino acid mutation is vs a data layer field capturing specific amino acid mutations detected in a protein sequence). Often, naive data specification developers use concept terms when it would be more appropriate to use information content entities (e.g., using "diagnosis" when a diagnosis is about a disease but is not itself the disease). These distinctions are not just semantic issues, but can cause logical clashes and errors when querying databases and extracting information from data management systems that use the ontology relationship axioms.

The above does not describe all possible data challenges, or even all that we have encountered, but rather highlights some of the most common barriers to harmonization and interoperability making them low-hanging fruit which could be the most impactful if addressed. To this end, below we suggest 10 simple rules for data standardization development that we have implemented in our own practice to help reduce these barriers.

## The Rules

To help users navigate the contextual data challenges described, we have outlined a series of 10 rules:

1. Reuse Existing Terms
2. Keep Concepts Simple, Specific & Consistent
3. Use Technical (not Colloquial) Words
4. Keep Terms as Universal as Possible
5. Avoid Abbreviations
6. Use Internationalized Resource Identifiers
7. Provide Definitions
8. Update and Deprecate
9. Use Tags to Avoid Ambiguity
10. Use Well-Maintained Ontologies

These rules are based on the assumption you have begun your standardization journey - that you have distilled out the unique concepts from your data descriptions, have an intended organizational structure (i.e., what values belong in which fields), and ideally have a strong understanding of how your concepts differentiate from one another (Figure 1).

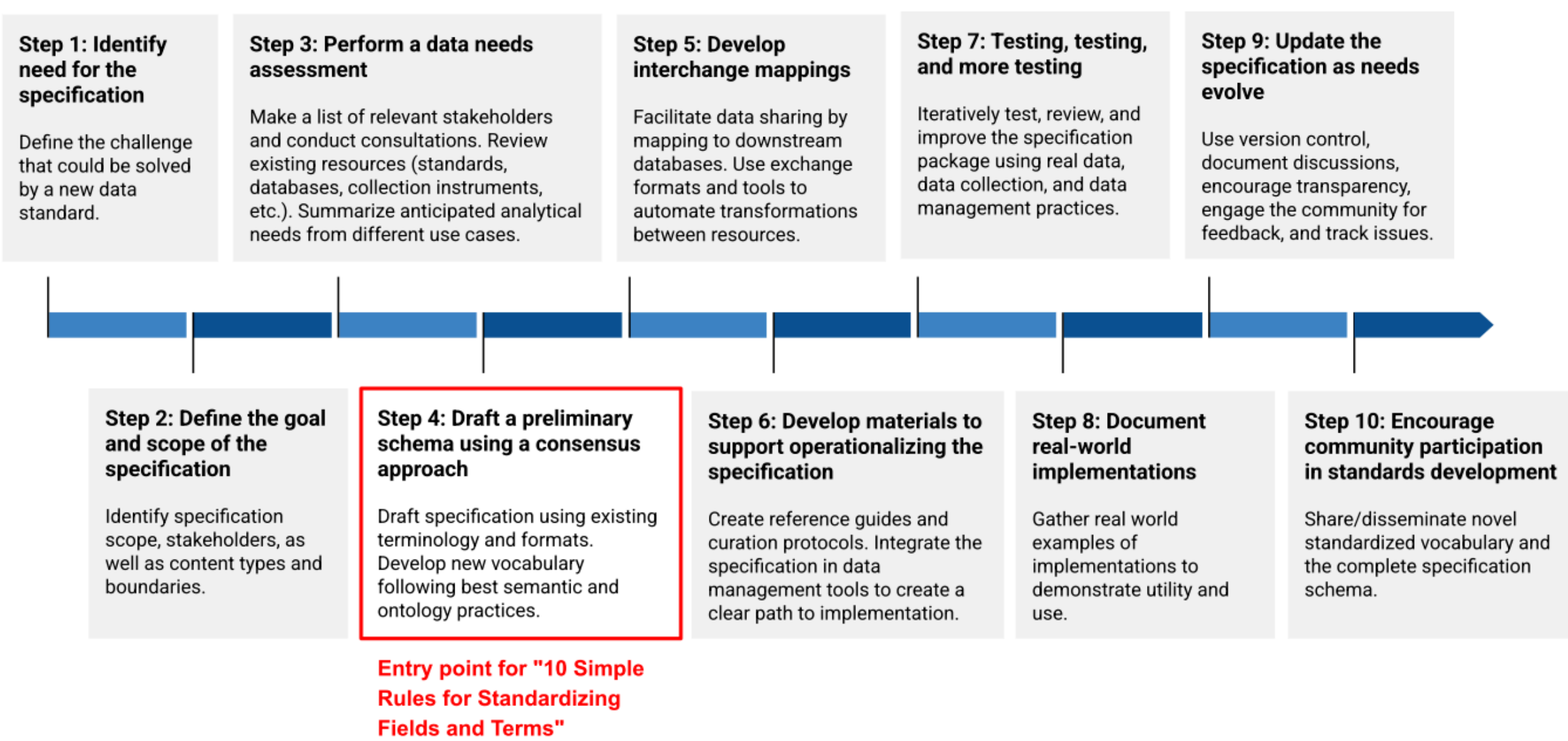

Figure 1: Overview of the ten-step method for ontology-based pathogen genomics standards development. There are many steps to developing a data specification which include 1) Identification of the need for a new specification, 2) Design specification goals and scope, 3) Perform a data needs assessment, 4) drafting and testing the new specification schema, 5) Developing interchange mappings, 6) Developing support and operationalizing materials, 7) Testing the specification package, 8) Documenting real-world implementations, 9) Updating and maintaining the specification, and 10) Dissemination and encouraging community participation. The entry point in the overall process referenced in this work is highlighted in red. *Note* Adapted from Griffiths *et al.* “10 Simple Rules for Developing Ontology-based Data Standards for Pathogen Genomics Contextual Data” [Unpublished manuscript][22].

## Rule 1: Reuse Existing Terms

First and foremost, do not “reinvent the wheel” - do not reinvent terms that are already addressed by existing ontology vocabulary. This rule aligns with the OBO Foundry principle 10 “commitment to collaboration” which advocates for reusing content to avoid work duplication and increase interoperability[3,23]. OBO Foundry ontology terms can be found using an ontology browser (Ontobee[6], Ontology Lookup Service (OLS)[5], etc.) or ontology tools such as the Protégé[8] editing environment. We recommend that beginners use an ontology browser until you become familiar with the ontologies that work best for you as they enable searching for terms across multiple domains. If you wish to use a non-ontology vocabulary for building your specification, we still recommend following these rules to help you evaluate the resource and its contents.

Here are some tips to help determine which terms are an appropriate match:

- Keep evaluating terms past the first search result, first is not necessarily the best.
- Try searching for synonyms, the ontology curators may have chosen a different label than you have.
- Look at the term definitions, they tell you more about the concept than the label.
- Look at the term's position within the ontology hierarchy, the upper and lower level concepts also inform the term’s meaning and intent.
- Check to see if the term is being reused across multiple ontologies, a reused term has been vetted by other ontology community members.
- Investigate the source ontology, consider if its domain of knowledge representation makes sense for your use case and whether it is actively maintained (Rule 10).

If you cannot find appropriate matches for your terms, you will likely need to create new terms and submit them to an appropriate existing ontology (you should get contributor credit that you can cite in grants and reports). If this is the case, proceed to Rules 2-10 to learn how to prepare standardized ontology terms for new term requests (usually accessible via an ontologies homepage or GitHub repository). Remember, you do not need to be perfect in a new term request, expert curators will help you refine your work, but doing your best to address these rules in advance will improve the likelihood and speed of acceptance, as well as the accuracy of the term for your data description needs.

**Example**

A data standards developer is looking to standardize terms for patient signs and symptoms, specifically looking for something along the lines of “cough” or “wet cough”. They navigate to ontobee.org and search for the keyword “cough”. A variety of results appear, including but not limited to “cough”, “frequency of coughing”, “how often cough”, “cough distress”, “cough drop”, “coughing up of blood”, and many more. Under an ontology labeled “HP” for “Human Phenotype” they find the term “productive cough” with the exact synonyms “wet cough” and “cough with mucus production”. Looking more closely at the term, they see “productive cough” is a subclass of “cough” that describes a variety of other coughs such as “nonproductive cough” and “chronic cough”. They decide to use both “productive cough” as well as “cough”, making a note that they may come back to this ontology branch for more cough terms. The benefit of using both classes is that the more specific “productive cough” term can be grouped with its parent class “cough” in analyses if necessary, but the data can still be standardized and captured at the appropriate level of detail in the lab management system.

## Rule 2: Keep Concepts Simple, Specific & Consistent

Remember the idiom "nobody reads the manual"? While this is not strictly true for everyone, data curation is a task that many people want to complete quickly so that they can move on to other priorities. As such, you need to keep in mind that many if not most of your users will not take the time to read all the information and guidance you provide, so the simpler, intuitive and less unambiguous you make a term, the more likely it is that your user will use it correctly. That being said, the desire to be concise should not be to the detriment of specificity. When you see the word "plasma" all alone, without additional information, what do you infer? Was it the "plasma" component of blood? Was it "plasma" as in the state of matter? Or perhaps it was the lipid and protein filled "plasma membrane" of cells? This potential for different understandings can guide individuals to multiple interpretations and data mismatching.

When conceptualizing your vocabulary, keep in mind what "thing"/"entity" you are describing and what it can be used for. Are you describing it or its application? Is it unique from similar terms within your descriptors and if yes, how so? Sort out the boundaries of material, processes, characteristics, and how they fit together to help you conclude the appropriate concept for your use case. It should tell us what something is, not what something is not, as the absence of something can be better conveyed through data queries within a database and during analysis (e.g., query "not" "diagnostic testing" rather than having a separate term for "not diagnostic testing" or "all meat except lamb")[24,25]. More specifically, it is best to avoid negative determiners (e.g., "no") or pronouns (e.g., "neither"). However, there may be commonly used negative nouns/adjectives where exceptions need to occur to appropriately describe a concept (e.g., "inorganic", "nonlinear transformation", "non-parametric test", etc.)[25]. Other naming conventions to keep in mind include using singular nouns unless the concept is plural by nature (e.g. "goggles") and single phrases only, do not use conjunctions (and) or disjunctions (or). Conjunctions are permitted in the cases of proper nouns (e.g. "Fisheries and Oceans Canada"), while disjunctions should be addressed by utilizing a broader label that encompasses the "or" (e.g. "poultry" rather than "chicken or turkey").

**Example**

When working on a controlled vocabulary for describing food-borne pathogen surveillance, the standards developer needs to differentiate between the different meanings of "harvest". So, they make the following set of terms for data descriptions: "harvested food material", "harvest season", "harvesting process", "harvesting machine", "harvest proportion", "harvest density", and many more. Even "harvest" terms not utilized for the data descriptions were considered to ensure that the terms applied did not inadvertently overlap with other meanings.

## Rule 3: Use Technical (not Colloquial) Words

We stated earlier the importance of using simple concepts with a clear label, but this should not be confused with a recommendation to use colloquial terminology. Concept descriptors should be formal, avoiding colloquialisms or jargon which are not readily understood by others outside of yourself, your organization, or your region. This form of vague, imprecise terminology is not useful for data sharing due to the potential for misunderstanding, and is prone to loose definitions or changes in meaning that may not persist with the dataset (i.e., anti future-proof). Well structured, technical vocabularies that are intelligible to scientists are also more likely to be amenable to querying and natural language processing.

**Example**

When working on a dataset which requires anatomical descriptors, a data collector indicates the location for which a sample/biopsy was taken as being the "belly". What exactly are they referring to in this case? Is it the "abdomen", the “stomach”, the “lower front part of the body”? Is it external or internal? This ambiguity not only makes it difficult to compare this data point with samples that use more specific data descriptors, it also impacts the conclusions a research scientist can make during their analysis. To prevent this, the specification template developer should provide a selection of standardized terms that encourage the right level of technical language use.

## Rule 4: Keep Terms as Universal as Possible

Hold on, were we not just saying to be specific? Now we are telling you to be non-specific? Well, yes and no. As mentioned previously, it’s important to be specific in the creation of your concept to appropriately describe the concept at hand and avoid ambiguity with other, similarly labeled terms. However, the key to this rule is that you should make a concept no more specific than necessary for your application. The addition of extraneous information usually reduces interoperability with other data descriptions, making a concept too narrow to include things that would have otherwise been comparable[24]. This rule applies to both labels and definitions.

**Example**

When developing contextual data descriptions for wastewater surveillance, a curator labeled “purpose of sequencing” as “wastewater purpose of sequencing”. The latter label is actually more narrow than the data descriptor described, there is no need to create a narrower term of “wastewater purpose of sequencing” when the original “purpose of sequencing” will suffice for the specification. Adding the narrower version artificially bloats the ontology and creates unnecessarily detailed terms that others will have to sort through. Additionally, when combining the described dataset with other data sources, it would result in an imperfect match with data described as “purpose of sequencing” when really they should be an exact match based on the available information.

## Rule 5: Avoid Abbreviations

While abbreviations may be extremely convenient, their usage in isolation increases the likelihood of miscommunication. Abbreviations may be limited in their application to a specific field of knowledge, or a specific region/agency/group applying this knowledge, which causes complications when data users without that knowledge need to understand what is being conveyed. An abbreviation may also overlap with the shortened form of another label, confusing users and resulting in exact data matches for completely different concepts. Within the medical field, approximately one-third of abbreviations were found to be ambiguous with more than one possible expansion[26]. So even if a system just uses the abbreviation for a user interface (UI) label but then the full label for export, there is still an increased risk that data entry personnel misinterpret the option they are selecting.

A concept should have the full expanded terminology as the preferred label (e.g., “Bronchoalveolar lavage”), with abbreviations (e.g., “BAL”) being applied as an additional annotation, which is useful for both search interfaces and text-mining applications. For data collection, it would be appropriate to append the abbreviation as an alternative/UI label (e.g., “Bronchoalveolar lavage (BAL)”), but the ontological primary label should instead include the abbreviation as an annotation. Exceptions are made for abbreviations that have become accepted as a word (e.g., “laser” as opposed to “light amplification by stimulated emission of radiation (LASER)”)[24].

**Example**

Data entry personnel are entering patient case data. Under a “tests performed” field, one individual lists “LFT” to describe “Liver Function Test”, while another uses “LFT” to describe “Lung Function Test”. A downstream analyst categorizes both data points as the same but does not know what “LFT” means and ends up having to drop them both for their analysis. Another analyst thinks both are referring to “Lung Function Test” and unintentionally misrepresents the results of their analysis. Having provided feedback to the systems manager, the data entry system restricts the field to a list value which includes both “Liver function test (LFT)” and “Lung function test (LFT)”. Now, users are not able to enter “LFT” alone into the data field, and once they begin typing “LFT” they are presented with both the “liver” and “lung” options to choose from and clarify their intended meaning. Downstream analysts are now able to differentiate and understand these values, using them to draw more nuanced conclusions from the dataset.

## Rule 6: Use Internationalized Resource Identifiers

Internationalized Resource Identifiers (IRIs) are URIs that have been standardized to a universal character set (Unicode/ISO 10646) but are mappable to URI schemes for backwards compatibility[27]. IRIs may be standalone references to a resource or Uniform Resource Locators (URLs) that identify a specific location for said resource. The OBO Foundry requires all ontology terms to have a unique IRI available as a standalone identifier and in the format of a persistent URL (PURL) that is permanent and redirectable in case the resource relocates[28]. Even if you are not specifically using OBO Foundry ontologies, the practice of using IRIs is extremely valuable. They facilitate the retrieval of resource information (e.g., annotation properties such as data mappings, data specifications, etc.) and are unique to a specific domain namespace, disambiguating terms with the same labels/synonyms for both users and computers, while also providing information to users on how a term may be interpreted based on the domain of knowledge the namespace addresses. Using IRIs also means you can use your own system labels (e.g. “Tachypnea (Rapid Breathing)” for “Tachypnea” http://purl.obolibrary.org/obo/HP_0002789) as long as the reference material and or export includes the IRIs and the IRIs are available for lookup and disambiguation. IRIs act as a “key” for switching to labels in other languages while also facilitating cross referencing identifiers in other controlled vocabularies and standards.

**Example**

A data analyst has found two instances of “plasma”, one assigned the IRI “NCIT:C13356” and the other “ENVO:01000798”. Using an ontology look-up service, they find “http://purl.obolibrary.org/obo/NCIT_C13356” links to a resource describing blood plasma, while “http://purl.obolibrary.org/obo/ENVO_01000798” links to a resource describing plasma as an environmental material state. Thanks to this information, the analyst can clearly differentiate between the inadvertently ambiguous labels. Additionally, the analysis knows that “NCIT” is the “National Cancer Institute Thesaurus” and “ENVO” is “The Environment Ontology”, so when they see IDs from these namespaces in the future they have a better idea of the domain of knowledge they are describing.

## Rule 7: Provide Definitions

Definitions are foundational in providing consistent communication of the intended meaning of a concept/term/field. Definitions are resources that help users within or outside of the described discipline use vocabulary appropriately. They are also useful for non-native language speakers who need clarification on vocabulary translations. Definitions should be straightforward, allowing users and curators to clearly understand the intended meaning[24] and be unique so as not to create unlinked, redundant vocabulary. The information in the definition should be true for all cases as all subclasses beneath inherit its meaning. Extraneous or subclass specific details should be left out, either included in comments/editor notes or as narrower classes where the information can be universally applied.

Ontological definitions have a consistent, logical structure that is helpful in maintaining consistency between ontologies, preventing common errors, and enabling translation into machine actionable data structures[3,24].

Ontological definitions should be singular and Aristotelian; formalized and normalized in the form "A [class] is a [parent class] that / which has [feature or differentia]", with additional (necessary) features/differentia added as list items to the sentence[3,24]. When defining a [class], the definition starts by stating the parent class under which it is situated in a hierarchy, so determining its domain and branch location will help with this process. Its hierarchical position anchors the term against more general classes; even if a class is polyhierarchical, usually only one parent class is referenced within the definition. The position within the hierarchy is one of the reasons ontological definitions are able to be so simplified, as terms also inherit the features/differentia listed in the definitions of parent classes.

Other things to keep in mind is that only one formal definition is necessary per term, each definition should be unique across terms (if they are not, then the terms should be combined via synonymy or further investigated to find differentia that disambiguate the two), use controlled vocabulary when referencing other vocabulary, are non-circular (e.g., where two concepts define themselves by referencing each other and thus do not provide clarifying information for either concept), and provide sources for definitions[24]. Additional information that should be made available, but not within the definition, are “example of usage” and user guidance “comment/editor note” annotations. While ontology/formal definitions have a prescribed structure, it is permissible to use more user-friendly definitions for specification reference guides as long as the simplified meaning still aligns with the technical meaning and does not inadvertently introduce inconsistencies with the ontological one. Always consider whether the information you wish to add may be more suitable as user guidance, especially if it is specific to your application.

**Example**

When developing metadata descriptions for food surveillance, a term curator defined data labeled “pasta” as “pasta food product” that is described as “...made from unleavened wheat flour dough and water…” in the reference materials. The curator has inadvertently made all references to “pasta” as more narrowly referencing “wheat pasta”. This is something another user may not realize when selecting the “pasta” term, inadvertently labeling gluten-free pasta as containing “wheat” - while another user who read the definition, may not have utilized the label because they realized it actually was not broadly describing “pasta” in general but rather a specific group of pasta in conflict with their “gluten-free pasta” dataset. This is remedied by making a new release of the terms, where “pasta food product” and its associated IRI are deprecated and replaced by a “wheat pasta food product” to more accurately reflect the original definition, and a new “pasta food product” is made that is defined as pasta “...made from unleavened flour dough and water…”.

## Rule 8: Update and Deprecate

In an ideal world our first attempt at solving a problem would result in something perfectly suited for both user needs and the problem at hand; however, the more we engage with the community, the more we come to recognize areas where the standard needs to adapt to better address different use cases. Standards need to evolve with best practices, data collection capacity/practices, changes in technology/methodology, and the research questions that need to be addressed by the collected data. To prevent a developing data standard from losing its capacity to standardize and harmonize, we should engage in the practice of ontological term deprecation and normalization.

Deprecation is when a term is deemed “obsolete”, usually with the intent of replacing it with a more appropriate term. When a field or value is removed from a data standard, users who are still catching up to the latest version may incidentally be using an older version and thus outdated terms. When we make a term "obsolete" we do not remove it from the ontology, the IRI/PURL remain, but "obsolete" is appended to the label and a new annotation is associated that redirects to the replacement term. Consequently, obsolete terms never leave a user stuck with an outdated, dead-end resource. Deprecation, and in fact all changes, in both ontologies and specifications should be tracked and version controlled. Database systems that are reusing ontologies have enough information from the "replaced by" deprecated term annotation to normalize to using the replacement terms in cases where a user inadvertently utilizes outdated versions. That being said, since 3rd party consumers of database content must be taken into consideration too; normalization can be performed using automated term replacement systems that have been manually mapped to the new terms.

**Example**

During the development of a data standard for describing respiratory illness symptoms, a list was created that included both "Tachypnea" and "Rapid Breathing" based on symptoms text mined from relevant research papers. Once users began engaging with the data standards, a medical expert recognized that "Rapid Breathing" is just the lay terminology for "Tachypnea", so users were picking between two terms that mean the same thing. Once this is reported, the ontology term created to describe "Rapid Breathing" is deprecated and replaced by "Tachypnea", and the "Tachypnea" term is given a synonym/layperm term annotation "Rapid Breathing". Now, when users input "Rapid Breathing", the system knows via the "term replacement", synonymy, or custom normalization process that the value should instead be "Tachypnea". Specifications using "Rapid Breathing" should preferably remove this term. If users prefer the layman's term, "Raping Breathing" could be replaced by including the synonymy in the "Tachypnea" term (i.e., "Tachypnea (Rapid Breathing)").

## Rule 9: Use Tags to Avoid Ambiguity

"Attribute tags" or "tags" are descriptors that provide additional information about a dataset or piece of data. All contextual data fields and terms could be viewed as attribute tags. Tagging datasets with informative attribute descriptions facilitates the appropriate identification and utilization and specific data types by surveillance practitioners and their computer systems (e.g., "purpose of sequencing" tags enable genomic epidemiologists to understand whether a sequence represents "baseline surveillance" vs "targeted surveillance"). There is a lot of information within a dataset, so descriptive tags are a good means of identifying dataset quality, utility, and accessibility. For example, quality control data tags can help identify low quality datasets, which are used in quality metrics for bioinformatics tool testing as well as educating trainees on how to think critically and identify common issues, and high quality data datasets for studies that may require a certain quality threshold before incorporating it into their work[18]. Additional data tags that describe the types of documentation data, characterization data, quantification data, and other more specific metadata, assist researchers in identifying useful datasets and Artificial Intelligence (AI) in training on what data elements are relevant to specific patterns of contextual data usage. Tags should avoid the use of boolean values, as there can be a variety of different data types described in a data tagging field, and if they end up separated from the original field container, the "TRUE/FALSE" or "YES/NO" will not have context for their original meaning. Additionally, tag what is present and what has been done as opposed to what is not present and what was not done as neither of the latter are informative and the list could be infinite.

**Example**

A dataset is being tagged using a series of questions such as "Was the primer specification deprecated?" "yes", "Were quality control issues identified?" "yes", "What level of quality control issues were identified?" "minor", and "High sequence quality threshold?" "no". Now all of these questions must be maintained in addition to their answers in order for the metadata tags to be utilized. To simplify this process, repeatedly used tags were reformatted and standardized, resulting in the dataset being tagged “primer specification deprecated”, “minor quality control issues identified”, and “low quality sequence”.

## Rule 10: Use Well-Maintained Ontologies

*Note: most of this advice also applies to non-ontological controlled vocabularies, so please consider it even if you are not engaging in ontology community practice for the development of your specification.*

Unfortunately not all ontologies are created equal. Ontologies tend to be built with a specific project in mind, even domain knowledge ontologies were initially built for the needs of a specific project that ultimately flavored its development. In order to appropriately prioritize ontologies for specification development work, consider the following:

**Is the ontology under active development?** You can check the most recent activity of an ontology by navigating to it on ontology lookup services or the version control website they use to manage their code (usually GitHub). Have they had any new releases in recent years? Are they responding to GitHub issue requests? If there hasn’t been activity in recent years it might not be active enough to process your term request.

**Is the ontology engaging in new term requests?** Ontology developers should be providing a process for users and other developers to submit new/change term requests. This is often provided using the GitHub “issues” feature or website/platform specific webforms, ideally with guidance on what information they would like users to provide with their requests.

**Is the ontology adhering to best practices?** The OBO Foundry operates on a set of guiding principles to ensure ontology interoperability and reduce redundancies. You can look to see if an ontology follows these principles by checking that they are listed in the OBO Library (http://obofoundry.org) and by consulting the OBO Foundry Dashboard (http://dashboard.obofoundry.org), a tool developed by the OBO Foundry technical working group to make it easier to perform high-level quality control checks. If you are not using an OBO Foundry ontology, you should be prioritizing vocabularies that have permanent IRIs and are ideally interoperable and/or mapped to other existing vocabularies.

**Is the ontology widely used?** Aside from when ontologies promote their usage in publications or on their web resources, it can be difficult to tell how reused ontology vocabularies are when reviewing them. One thing to help you identify reusability when searching terms is to check to see what other ontologies have imported the term you are looking at, the greater the reuse the more vetted the term. On OntoBee (http://ontobee.org), imports are indicated as bullet points underneath a term search result, while EMBL-EBI OLS (http://www.ebi.ac.uk/ols/index) indicates imports underneath a term search result with the “Also appears in” header.

**Is the ontology providing definitions, sources, and other annotations?** As described in Rule 7, definitions are an important function of ontology, or any controlled vocabulary, terms and thus it is important that the ontologies you use engage in the practice of providing definitions and their respective sources (when not completely authored by the term creator). Bonus points if they are also including annotations indicating synonymy, examples of usage, and axioms that break down complex concepts into their respective parts.

If you’ve found an ontology you are having difficulty evaluating, ask an ontologist for their feedback on whether it would be relevant to your work. It is also important to keep in mind that sometimes ontologies struggle with a lack of funding and consequently curation/developer labour. Just because an ontology is not being actively managed or is being slow to respond does not mean it is of poor quality, but it does mean it may have difficulty meeting your specification needs on your timeline. If you've requested a term in an ontology and it is not being addressed, it is appropriate to withdraw your request and redirect it elsewhere. For a deeper dive into the subject, check out "Ten Simple Rules for Selecting a Bio-ontology"[29].

**Example**

A data curator is curating a list of symptoms and is trying to find an appropriate match for "Tachypnea". When you search it on an ontology look-up service you find several hits. They find a few exact matches, but only some have definitions, and one hit also appears to be reused in four other ontologies. Investigating further they see the term also indicates the definition source, a layperson term label, cross references that map to codes used to describe this concept in other databases, additional information expanding on the term's meaning, and equivalent axioms that break the concept down into sub concepts that could be useful in database querying. When checking the ontology version information, they see it has had a new release within the last year and their GitHub is actively updating code and addressing issue requests. Not only have they found their term for "Tachypnea", they have found this ontology to be well executed and will keep it in mind for future symptom term matches and new term requests.

## Conclusion

Standards are a critical part of our pathogen-genomics data ecosystem because they enable faster and more efficient data management, harmonization, integration and analysis. Genomic sequence data has many different uses, and time spent on applying standards to contextual data is an investment that pays off in the present and also increases the likelihood of reusability in the future. Interoperability of data standards does not manifest out of nowhere, but rather must be engineered based on community consensus, community resources, and best practices. Applying clear, well-articulated, harmonized standards development practices in the present reduces burdens on data curators, pathogen-genomic surveillance practitioners, and researchers in the long-term, while also decreasing the likelihood of data loss from ambiguous or incompatible data values. Yet, there is a lack of knowledge translation and accessible skill training in applying these practices to standards creation, which is why we have drafted a minimum set of 10 rules to help data harmonization efforts. While this work comes from our experience in genomic public-health surveillance we strongly believe these can be applied to other domains of knowledge, facilitating the development of FAIR, linked data standards that generate understandable and readily reusable data.